\documentclass{aa}
\usepackage{epsf}

\usepackage{graphicx}
\usepackage{aabib99}

\newcommand{\La}{\mbox{${\rm Ly\alpha}$}}
\newcommand{\Line}[3]{\Ion{#1}{#2}\,$\lambda$\,#3}

\newcommand{\Ion}[2]{#1{\,\scriptsize #2}}
\newcommand{\Nh}{\mbox{$N_{\rm H}$}}
\newcommand{\Rwd}{\mbox{$R_{\rm wd}$}}
\newcommand{\Mwd}{\mbox{$M_{\rm wd}$}}
\newcommand{\Twd}{\mbox{$T_{\rm wd}$}}
\newcommand{\Mc}{\mbox{$M_{\rm c}$}}
\newcommand{\Tc}{\mbox{$T_{\rm c}$}}
\newcommand{\Rc}{\mbox{$R_{\rm c}$}}
\newcommand{\Tspot}{\mbox{$T_{\rm spot}$}}
\newcommand{\Lspot}{\mbox{$L_{\rm spot}$}}
\newcommand{\Teff}{\mbox{$T_{\rm eff}$}}
\newcommand{\Msun}{\mbox{$M_{\odot}$}}
\newcommand{\Porb}{\mbox{$P_{\rm orb}$}}
\newcommand{\es}{\mbox{$\rm ergs\;s^{-1}$}}
\newcommand{\msy}{\mbox{$\rm \Msun\,yr^{-1}$}}
\begin{document}

\thesaurus{06      
          (02.01.2 
           08.02.1 
           08.09.2 
           13.25.5 
          )}
\title{RX\,J1313.2$-$3259, a missing link in CV evolution?
\thanks{Based on observations made with the International Ultraviolet
Explorer}}

\author{
   B.T. G\"ansicke\inst{1},
   K. Beuermann\inst{1},
   D. de Martino\inst{2}
   H.-C. Thomas\inst{3}}

\offprints{B.T.G\"ansicke, boris@uni-sw.gwdg.de}

\institute{
Universit\"ats-Sternwarte, Geismarlandstr. 11, D-37083 G\"ottingen, Germany
   \and
Osservatorio di Capodimonte, Via Moiariello 16, I-80131 Napoli, Italy
   \and
MPI f\"ur Astrophysik, Karl-Schwarzschild Str. 1, D-85470 Garching, Germany}

\date{Received 5 October 1999~/~Accepted 25 October 1999 }

\authorrunning{B.T. G\"ansicke et al.}

\maketitle 

\begin{abstract}
We present low-state IUE spectroscopy of the ROSAT-discovered polar
RX\,J1313.2$-$3259. The SWP spectrum displays a broad \La\ absorption
profile, which can be fitted with a two-temperature model of a white
dwarf of $\Twd=15\,000$\,K with a hot spot of $\Tspot=34\,000$\,K
which covers $f\sim0.01$ of the white dwarf surface.
The white dwarf temperature is atypically low for the long orbital
period (4.18\,h) of RX\,J1313.2$-$3259. 
This low temperature implies either that the system is a young CV in
the process of switching on mass transfer
or that it is an older CV found in a prolonged state of low accretion
rate, much below that predicted by standard evolution theory.  In the
first case, we can put a lower limit on the life time as pre-CV of
$10^8$\,yrs. In the second case, the good agreement of the white dwarf
temperature with that expected from compressional heating suggests
that the system has experienced the current low accretion rate for an
extended period $>10^4$\,yrs.  A possible explanation for the low
accretion rate is that RX\,J1313.2$-$3259 is a hibernating post nova and
observational tests are suggested.

\end{abstract}

\section{Introduction}
Most fundamental physical stellar parameters of field white dwarfs,
such as effective temperature, surface gravity, and magnetic field
strength can directly be measured with high precision from
spectroscopic observations. Assuming a mass-radius relation, both mass
and radius may be inferred independently of the distance.
Determining these properties also for the accreting white dwarfs in
cataclysmic variables (CVs) is a relatively new research field,
essential not only for testing stellar (binary) evolution theory,
but for understanding the physics of accretion in this whole class
of binaries.

The last years saw a rapid growth of identified polars, CVs containing
a synchronously rotating magnetic white dwarf. Despite the large
number of know systems ($\ga60$) rather little is known about the
temperatures of the accreting white dwarfs in these systems. The main
reasons for this scarcity are twofold.
(a) In the easily accessible optical wavelength band, the white dwarf
photospheric emission is often diluted by cyclotron radiation from the
accretion column below the stand-off shock, by emission from the
secondary star, and by light from the accretion stream.  Even when the
accretion switches off almost totally and the white dwarf becomes a
significant source of the optical flux (e.g.  \nocite{schwopeetal93-1}
Schwope et al. 1993), the complex structure of the Zeeman--splitted
Balmer lines and remnant cyclotron emission complicate a reliable
temperature determination.
(b) At ultraviolet wavelengths the white dwarf entirely dominates the
emission of the system during the low state and may be a significant
source even during the high state. However, the faintness of most polars
requires time-consuming space based observations
(e.g. \nocite{stockmanetal94-1} Stockman et al. 1994).

\section{Observations}
IUE observations of RX\,J1313.2$-$3259 (henceforth RX\,J1313) were
carried out in March, 1996.  One SWP (1150$-$1980\,\AA) and one LWP
(1950$-$3200\,\AA) low resolution spectrum were obtained on March 2
and March 6, respectively (Table\,1). The LWP image was taken prior to
the failure of Gyro\,\#5, read-out of the image had to await that
control over the spacecraft was re-established. Both observations were
taken through the large aperture, resulting in a spectral resolution
of $\approx6$\,\AA. Because of the faintness of RX\,J1313, the
exposure time of the SWP spectrum was chosen roughly equal to the
orbital period. The spectra have been processed through the
IUE\,NEWSIPS pipeline, yielding flux and wavelength calibrated
spectra.

The SWP spectrum is shown in Fig\,\ref{f-swp}. It is a blue continuum
with a flux decline below $\approx1400$\,\AA. Due to the long exposure
time, the spectrum is strongly affected by cosmic ray hits. Some
emission of \Line{C}{IV}{1550} and \Line{He}{II}{1640} may be present
in the spectrum of RX\,J1313, but from the present data no secure
detection of line emission can be claimed.

\section{Analysis and Results}

\begin{figure*}
\parbox{8.8cm}{\includegraphics[angle=270,width=8.8cm]{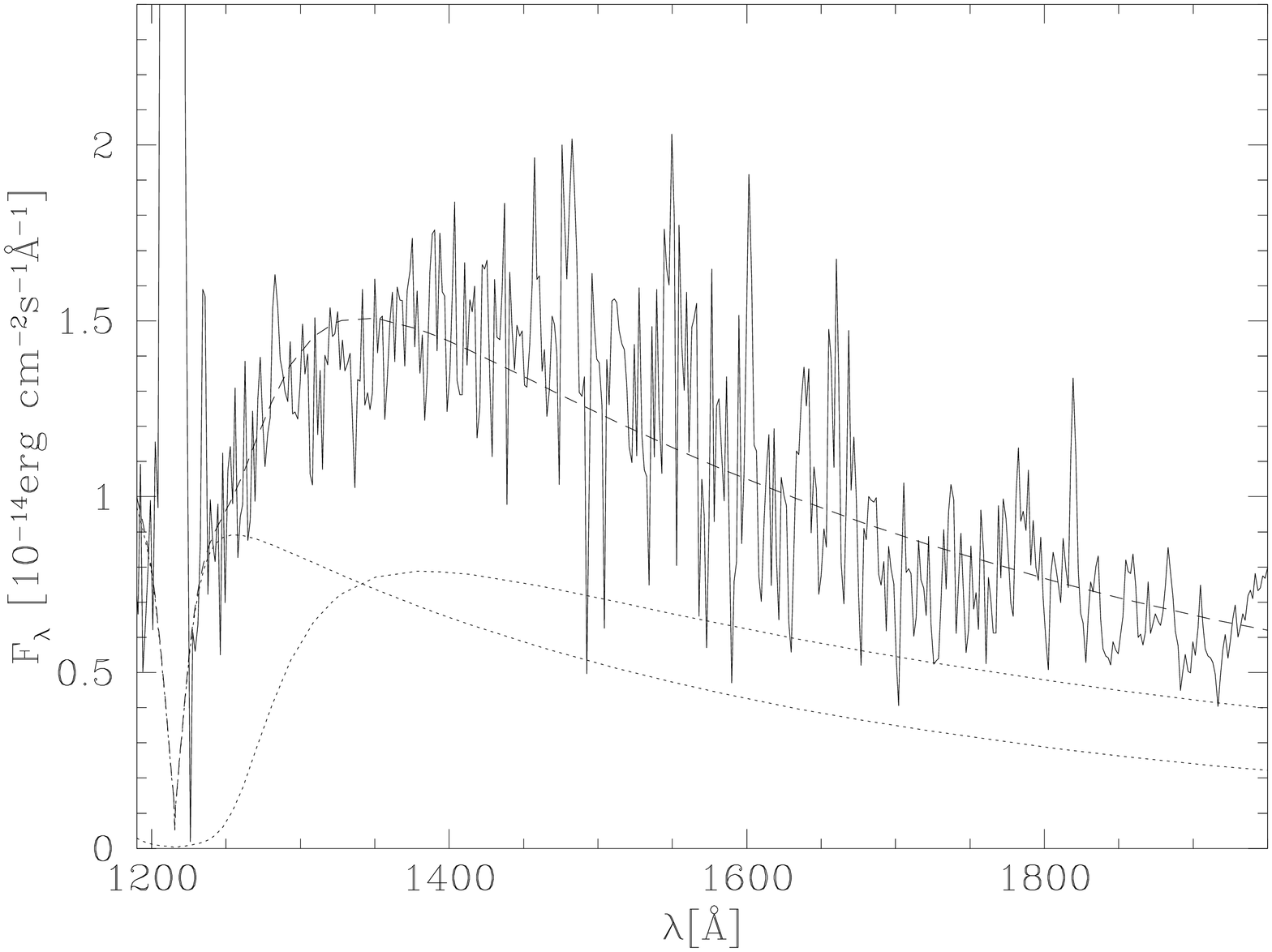}}
\hfill
\parbox{8.8cm}{\includegraphics[angle=270,width=8.8cm]{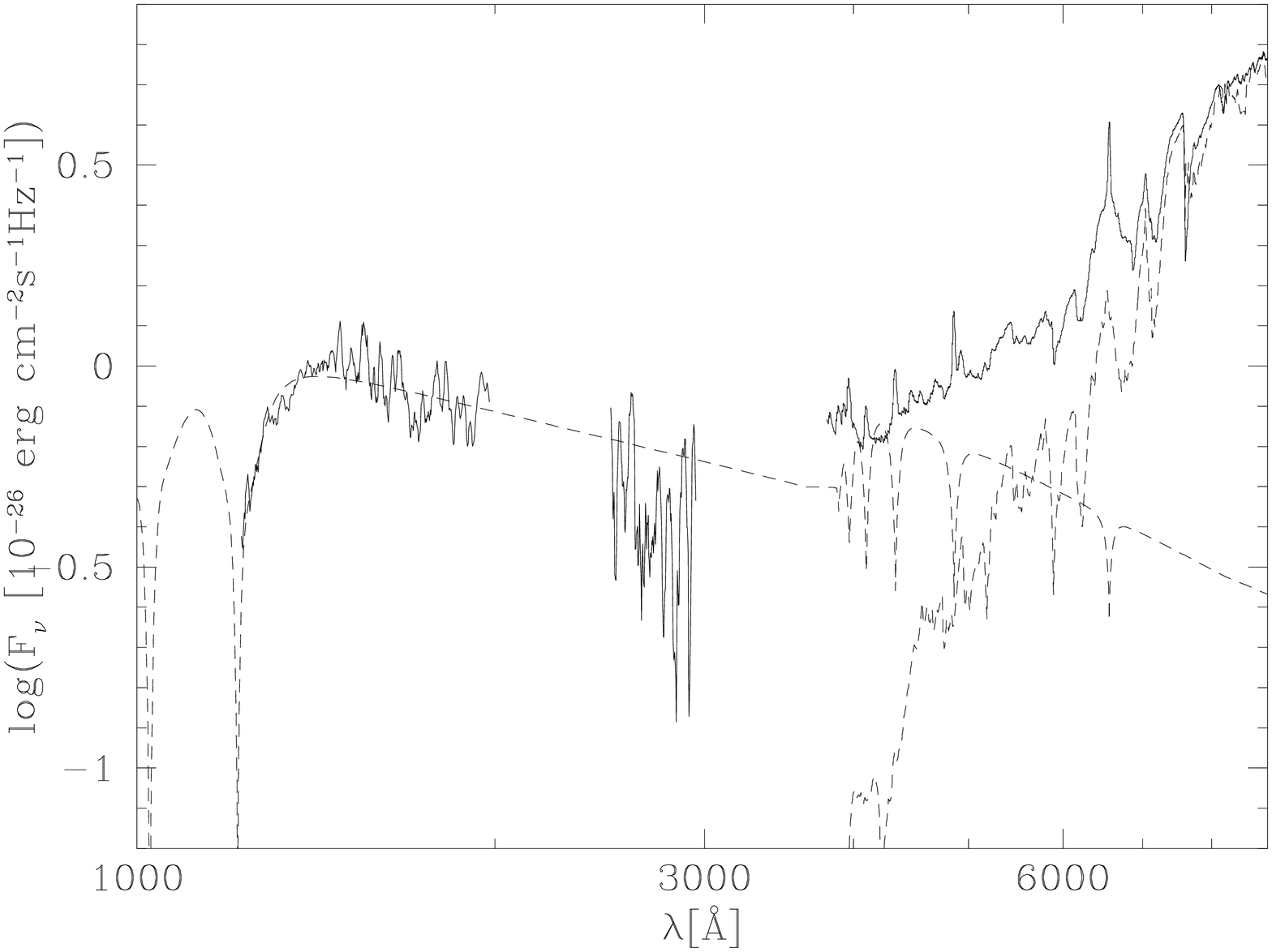}}

\parbox[t]{8.8cm}{\caption[]{\label{f-swp} IUE\,SWP low state spectrum
of RX\,J1313. Plotted as a dashed line is the best-fit two
temperature model of a white dwarf of $\Twd=15\,000$\,K with a hot
spot of $\Twd=34\,000$\,K covering $f\sim0.01$ of the white dwarf
surface. The dotted curves show the individual contributions of the
two components.}}
\hfill
\parbox[t]{8.8cm}{\caption[]{\label{f-overall} Non-simultaneous
IUE\,SWP\,/\,LWP and optical low state spectra of RX\,J1313.
Plotted as a dashed line is  the best-fit two temperature model from
Fig.\,1 and an observed M3 dwarf. Note that only the SWP data were
used for the fit.}}
\end{figure*}

The absence/weakness of emission lines strongly indicates that the IUE
observations were taken during a period of very low accretion
activity. The broad flux turnover below $\approx1400$\,\AA\ is
reminiscent of the photospheric \La\ absorption observed during low
states, e.g. in AM\,Her \cite{szkodyetal82-1,gaensickeetal95-1} or
DP\,Leo \cite{stockmanetal94-1}. Our first approach was, thus, to fit
the SWP data with non-magnetic pure hydrogen white dwarf model spectra
\cite{gaensickeetal95-1}. However, none of the models could
satisfyingly describe the observed spectrum. While the continuum
requires a rather low temperature, $\approx15\,000$\,K, the steep slope
in the narrow core of the \La\ absorption ($1220-1300$\,\AA) is in
disagreement with the very broad \La\ line of such low-temperature
models.

\begin{table}[t]
\caption[]{IUE observations of RX\,J1313$-$32. Listed are the IUE
frame numbers, the observation dates, and the exposure times.}
\begin{flushleft}
\begin{tabular}{lrr}
\hline\noalign{\smallskip}
Image\,No. & Exp. start (UT) & Exp. time (sec) \\
\noalign{\smallskip}\hline\noalign{\smallskip}
SWP56879L & 02\,Mar\,1996\,08:01:49 & 13800 \\ 
LWP32069L & 06\,Mar\,1996\,18:20:31 & 2100 \\
\noalign{\smallskip}\hline\noalign{\smallskip}
\end{tabular}
\end{flushleft}
\vspace{-5mm}
\end{table}

The analysis of low-state ultraviolet spectroscopy of other polars
taught us that the white dwarfs often have a non-uniform temperature
distribution over their surface \cite{gaensicke98-2,stockmanetal94-1},
possibly due to heating by low-level accretion
\cite{gaensickeetal95-1}.
We, therefore, fitted the IUE data of RX\,J1313 with a two-temperature
model, using again our non-magnetic pure hydrogen model spectra and
leaving four free parameters; the temperatures and scaling factors of
both components. The best fit is achieved by a white dwarf with a
``base'' temperature of $\Twd=15\,000$\,K and a ``spot'' temperature
of $\Tspot=34\,000$\,K (Fig.\,\ref{f-swp}). For a distance
$d=200$\,pc, as derived by Thomas et al. \cite*{thomasetal99-1}, the
white dwarf radius resulting from the scaling factors is
$\Rwd=1.1\times10^9$\,cm. Assuming the Hamada-Salpeter (1961)
\nocite{hamada+salpeter61-1} mass-radius relation for a carbon core,
the corresponding mass is $\sim0.4$\,\Msun, which is consistent with
the mass derived by Thomas et al. \cite*{thomasetal99-1}.

Because the IUE/SWP observation represents the orbital mean of the
ultraviolet emission of RX\,J1313, the spot size cannot be directly
estimated. Assuming that the ultraviolet-bright spot shows a similar
variation as the X-ray spot observed with ROSAT
\cite{thomasetal99-1}, we estimate a fractional area
$f\sim0.01$. For a somewhat larger spot, the temperature would be
correspondingly lower.

Fig.\,\ref{f-overall} shows the IUE SWP and LWP spectra along with an
average optical low state spectrum, as well as the two-component
model.  The flux of the LWP spectrum is somewhat lower than predicted
by the model, which could be due either to heavy underexposure
(Table\,1) or to the fact that the LWP spectrum covers only
$\approx0.14$ of the orbital period, possibly resulting in a lower
spot-contribution than in the orbital-averaged SWP spectrum, or both.
The agreement of the model spectra with observed optical flux is
reasonably good, especially when considering that only the
1225--1900\,\AA\ range was used for the fit and that the ultraviolet
and optical spectra were taken at different epochs. The summed
spectrum of the white dwarf model and a red dwarf matching the red end
of the RX\,J1313 spectrum has $V=16.4$, which is in agreement with the
observed low-state magnitude of the system
\cite{thomasetal99-1}. During the low state, the optical and
ultraviolet emission of RX\,J1313 is, hence, dominated by its two
stellar components.

For completeness, we mention that an additional possible source of
\La\ absorption is the interstellar gas. We computed the interstellar
\La\ profile for the absorption column derived from the X-ray data,
$\Nh=9\times10^{19}\mathrm{cm}^{-20}$ \cite{thomasetal99-1}. The width
of this line is smaller than the geocoronal emission in the SWP
spectrum. Thus, interstellar absorption cannot explain the narrow
``core'' observed in the IUE spectrum.

\section{A note on the use of non-magnetic model spectra}
A major uncertainty in the computation of realistic hydrogen line
profiles in magnetic atmospheres is the treatment of the Stark
broadening in the presence of a magnetic field. The Stark broadening
of the individual Zeeman components is smaller than that of the entire
transition in the non-magnetic case, but no detailed calculations are
available. This uncertainty can be taken into account by treating 
the amount of the Stark broadening as a free parameter in the model
atmosphere calculation and calibrating it with observations
\cite{jordan92-1}. For \La, this approach is, however, difficult. On
one hand, there are only few single magnetic white dwarfs for which
good ultraviolet spectroscopy has been obtained. On the other hand,
the three Zeeman components of \La\ become visible as individual
absorption features only for fields $B\ga100$\,MG. For lower field
strengths the \La\ profile is still dominated by the Stark effect and
the Zeeman shifts introduce only an additional broadening which is,
again, difficult to quantify.

An additional problem in the computation of synthetic \La\ profiles
arises for low-temperature atmospheres ($\Teff\la20\,000$\,K).  In
ultraviolet observations of non-magnetic white dwarfs in this
temperature range, quasi-molecular absorption of $H_2^+$ and $H_2$
produces strong absorption features at $\approx1400$\,\AA\ and
$\approx1600$\,\AA, respectively, which are overlayed on the red wing
of \La\ \cite{koesteretal85-1}. Calculations of these transitions in
the presence of a strong magnetic field have not yet been approached.
We have retrieved the IUE spectra available for all magnetic white
dwarfs listed by Jordan \cite*{jordan96-1}, and find that in at best
two of them the $H_2^+$ feature can be identified (BPM\,25114,
$B\approx36$\,MG and KUV\,23162$-$1230, $B\approx56$\,MG).  Also, none
of the accreting magnetic white dwarfs in polars with
$\Teff\la20\,000$\,K observed in the ultraviolet display noticeable
$H_2^+$ absorption \cite{gaensicke97-1}. From Fig.\,\ref{f-swp} it is
apparent that also the spectrum of RX\,J1313 is devoid of noticeable
absorption at 1400\,\AA\ and 1600\,\AA. In summary, observations
indicate that the $H_2^+$ and $H_2$ quasi-molecular absorption lines
may be weaker in a strongly magnetic atmosphere than in a non-magnetic
one.

Assuming a magnetic field strength of $B=56$\,MG for RX\,J1313, as
derived by Thomas et al. \cite*{thomasetal99-1} from the cyclotron
emission, the expected shift of the $\sigma^+,\sigma^-$ components is
$\pm34$\,\AA, causing the centre of the $\sigma^+$ component to
coincide with the steepest slope of the \La\ profile.
While the Zeeman effect may broaden the observed \La\ profile, the
reduced Stark broadening will cause an opposite effect.  We estimate
that the use of non-magnetic model spectra in the analysis of the \La\
profile may cause a temperature error of a about $\pm1000$\,K. 

We conclude that the theoretical uncertainties in the Stark broadening
do presently not warrant the use of magnetic model spectra.  The
narrow core in the broad \La\ absorption observed in RX\,J1313 cannot
be produced by magnetic effects supporting our interpretation of a
rather cool white dwarf with a localized hot region.

\section{Discussion}

It is well established that the white dwarfs in CVs tend to be hotter
than single white dwarfs. This observational result suggests that
accretional heating takes place in addition to the secular core
cooling of the white dwarfs in CVs
(e.g. \nocite{sion91-1,sion99-1}Sion 1991,1999).
Furthermore, the white dwarfs in CVs below the period gap are on
average cooler than those in CVs above the gap (G\"ansicke 1997,1998;
Sion 1991,1999)\nocite{gaensicke97-1,gaensicke98-2,sion91-1,sion99-1}.
A combination of two effects is thought to be responsible for this
difference: (i) the average age of the short-period CVs below the
period gap is about an order of magnitude larger than that of CVs
above the gap \cite{kolb+stehle96-1} and core cooling of their white
dwarfs has progressed correspondingly; (ii) the average accretion rate
in short-period CVs is about an order of magnitude lower than in
long-period CVs \cite{king88-1}, resulting in reduced accretional
heating. Warner \cite*{warner95-1} shows -- admittedly only for a
small sample of CVs -- that the expected correlation between accretion
rate and white dwarf temperature does, in fact, exist.

RX\,J1313 is the polar with the fourth-longest period. It is,
therefore, expected to be rather young, to experience a comparatively
high time-averaged accretion rate, and to have a correspondingly hot
white dwarf. Contrary to these expectations, however, it harbours the
coldest white dwarf of all the CVs above the period gap.  In fact, the
temperature of the white dwarf in RX\,J1313 is comparable to the
average white dwarf temperature in short-period CVs.
We suggest two possible scenarii which can explain the atypically low
white dwarf temperature.

(a) RX\,J1313 has only recently developed from a detached
pre-cataclysmic binary into the semi-detached state. Mass transfer is
in the process of turning on and substantial heating of the white
dwarf has not yet taken place. In this case, the observed effective
temperature of the white dwarf allows to estimate a lower limit on the
cooling age and, thereby, on the time elapsed since the system emerged
from the common envelope. The time scale for the turn-on of the mass
transfer is $\sim10^4$\,yrs, which is short compared to the
$\sim10^8$\,yrs that a CV spends above the gap \cite{ritter88-1}. The
probability of finding a CV in this stage of its evolution is rather
small, but non-zero.

(b) RX\,J1313 is a ``normal'' long-period CV, but has more
recently experienced a low accretion rate for a sufficiently long time
interval ($>10^4$\,yrs) which allowed its white dwarf to cool down to
its current temperature. Long-term ($\tau \ga 10^4$\,yrs) fluctuations
of the accretion rate about the secular mean predicted from angular
momentum loss by magnetic braking \cite{king88-1} are consistent with
the large range of observed accretion rates at a given orbital period
\cite{patterson84-1,warner87-1}. Two possible explanation for these
fluctuations have been suggested.
(b1) A limit-cycle in the secondary's radius driven by
irradiation from the hot primary (King et
al. 1995,1996\nocite{kingetal95-1,kingetal96-1}) which causes a
corresponding variation in the mass transfer rate.
(b2) CVs possibly enter a prolonged phase of low (or zero)
$\dot M$ following a classical nova eruption, referred to as
hibernation \cite{sharaetal86-1}. RX\,J1313 may be such a hibernating
CV. However, the low temperature of the white dwarf in RX\,J1313
argues against a very recent nova explosion. In V1500\,Cyg, the white
dwarf cooled from the nuclear burning regime, i.e. several $10^5$\,K
in 1975 to $\sim95\,000$\,K in 1992 \cite{schmidtetal95-1}. On the
theoretical side, Prialnik (1986) shows in a simulation of a
1.25\,\Msun\ classical nova that the white dwarf reaches its minimum
temperature $\sim8000$\,yrs after the nova explosion.

Two observational tests could help to decide whether RX\,J1313 is a
rather ``fresh'' post-nova with its secondary only marginally filling
its Roche-lobe:
(1) A nova eruption may break synchronization, as observed in
V1500\,Cyg \cite{stockmanetal88-1,schmidt+stockman91-1}, causing the
orbit to widen and the secondary to retreat from its Roche lobe. The
resynchronization of the white dwarf spin with the orbital period
occurs in V1500\,Cyg apparently on a time scale of a few hundred years
\cite{schmidtetal95-1}. More accurate ephemerides of RX\,J1313 than
presently available (Thomas et al. 1999) are necessary to test for a
small remnant asynchronism of the white dwarf spin.
(2) The nova eruption may contaminate the binary system with material
processed in the thermonuclear event, resulting in deviations from the
typical population\,I abundances found in most CVs
\cite{marks+sarna98-1}.  Anomalous ultraviolet emission line ratios
similar to those observed in post-novae have been found in the
asynchronous polar BY\,Cam \cite{bonnet-bidaud+mouchet87-1}. The white
dwarf in BY\,Cam has $\Twd=20\,000$\,K \cite{gaensicke97-1}, which is
in rough in agreement with the temperature expected a few 1000 years
after a nova explosion. BY\,Cam contains also a slightly
asynchronously rotating white dwarf, which leaves the question of the
expected time scale for resynchronization somewhat unsettled.
High-state ultraviolet observations of the CNO lines in RX\,J1313 do
not exist so far and will serve to test the post-nova hypothesis. If
no evidence for a nova event should be found, RX\,J1313 is either very
young as a CV or experiences a prolonged low state in some kind of
mass transfer cycle.

Before we discuss the temperature of the white dwarf photosphere,
\Teff \,=\,15\,000\,K, we comment on the ``warm'' ultraviolet-bright
spot with \Teff $\,\simeq\,34\,000$\,K. RX\,J1313 is yet another polar
in which the white dwarf appears to have a non-uniform temperature
distribution. Other examples are AM\,Her \cite{gaensickeetal95-1},
DP\,Leo \cite{stockmanetal94-1}, or QS\,Tel \cite{demartinoetal98-1}.
These hot spots are best explained by the localized irradiation of the
photosphere with cyclotron and X-ray photons from the accretion funnel
which is continuously fed at a low rate. In the case of RX\,J1313, we
estimate the luminosity of the ultraviolet-bright spot to be
$\Lspot\simeq 10^{31}$\,\es, corresponding to an accretion rate of
$\dot M >3\times10^{-12}$\,\msy, which is consistent with the
low-state accretion rate, $\dot M \simeq 6\times 10^{-12}$\,\msy,
derived by Thomas et al. \cite*{thomasetal99-1}.

We now discuss accretion heating of the white dwarf in RX\,J1313.  As
shown by Giannone \& Weigert \cite*{giannone+weigert67-1} and by Sion
\cite*{sion95-1} this is an inherently time-dependent
process. Accretion compresses the outer non-degenerate layers of the
white dwarf which heat up approximately adiabatically if the accretion
rate is high. The core suffers some compression, too, which heats
primarily the non-degenerate ions.  For intermittent accretion, the
thermal inertia of the deep heating produces a time delay which causes
an enhanced luninosity long after accretion stopped.  For very low
accretion rates, $\dot M \sim 10^{-11}$\,\msy, on the other hand,
prolonged accretion may lead to a quasi-stationary state in which the
energy loss balances compressional heating \cite{giannone+weigert67-1}
and the temperature profile of the envelope remains stationary.

The simplest way to view compressional heating is to consider the
energy released when the accreted mass is added. Since the envelope
mass is small and represents a practically constant fraction of the
white dwarf mass, $\dot M$ increases the mass of the degenerate core
of mass \Mc$\simeq $\Mwd, radius \Rc, and temperature \Tc. The energy
released per unit time is G\Mwd$\dot M$(1/\Rc -- 1/\Rwd) of which some
fraction feeds the initial degeneracy of the electrons reaching
\Rc. Apart from a factor of order unity, this energy equals the work
performed by compression, $P\,{\rm d}(1/\rho$)/dt with $P$ the
pressure and $\rho$ the density, integrated over the envelope.  Note
that this energy release is different from that freed at the surface,
which equals G\Mwd$\dot M$/\Rwd\ in an AM Her star, and represents the
additional energy released by compression of the envelope of the white
dwarf between radii \Rwd\ and \Rc.  If compression is adiabatic, the
work performed is used to increase the internal energy of the gas, as
prescribed by the first law of thermodynamics. In the isothermal case,
the released energy would completely appear as radiative loss. We
consider here the case of slow compression and assume that the energy
released by accretion at a rate $\dot M$ equals the increment in
luminosity
\begin{equation}
L_{\rm acc} = \eta \,{\rm G}\Mwd \dot
M\left(\frac{1}{\Rc} - \frac{1}{\Rwd}\right) 
\end{equation}
where G is the gravitational constant, 
and we estimate that $\eta$ is between 0.5 and 1.0. Core heating is a
minor effect and adds only $\sim 10$\,\% to $L_{\rm acc}$. Hence, the
compressional energy is primarily released in the envelope, at an
approximately constant rate per radius interval.

In equilibrium, accretion at a rate $\dot M$ can maintain an effective
temperature \Teff\ of the white dwarf defined by \mbox{$L = 4\pi
\Rwd^2\sigma\Teff^4 = L_{\rm acc}$}, even if the white
dwarf had cooled to a substantially lower temperature before the
onset of accretion. 

In their discovery paper, Thomas et al. (1999) derive a mass of the
white dwarf in RX\,J1313 of \Mwd\,$\simeq~0.40$\Msun\ and a secondary
mass of $M_2 \simeq~0.45$\Msun, with uncertainties of about
0.10\Msun. There was some concern about the mass ratio which should be
$M_2$/\Mwd$<0.7$ for stable mass transfer. For definiteness, we assume
here \Mwd\,$\simeq~0.5$\Msun.
Thomas et al. (1999) also derived a mass accretion rate which is very low
compared to other long period CVs. They observed the system over seven
years and found that it hovers most of the time at low accretion
luminosities corresponding to $\dot M \sim 10^{-11}$\,\msy. Only twice
was the system found in an intermediate state with an accretion rate
of $\sim 10^{-10}$\,\msy, during the ROSAT All-Sky-Survey and in a
subsequent optical follow-up observation in February 1993. It was never
observed at an accretion rate of $\sim 10^{-9}$\,\msy, the typical
value of CVs with \Porb = 4--5 h \cite{patterson84-1}. To be sure, the
derived accretion rates depend (i) on the adopted white dwarf mass and
(ii) on the soft X-ray temperature and the bolometric fluxes of the
quasi-blackbody source, and the quoted rates are probably uncertain by
a factor of $\sim 2$.

A white dwarf of 0.5\,\Msun\ has a core radius \Rc = $9.3\times
10^8$\,cm and a radius \Rwd$\,\simeq 1.05\times 10^9$\,cm at \Teff =
15\,000\,K. For these parameters, we find an equilibrium temperature
from compressional heating alone of \Teff\ $\simeq 16\,400\,(\eta\dot
M_{-10})^{1/4}$~K, where $\dot M_{-10}$~K is the accretion rate in
units of $10^{-10}$\,\msy. The observed temperature of 15\,000\,K can
be maintained by an accretion rate of $7\times 10^{-11}$\,\msy\ for
$\eta \simeq 1$, which is within the observed range of mass transfer
rates. Since the internal energy source of the white dwarf will
contribute to the observed luminosity, the actual accretion rate
required to maintain the photosphere at 15\,000 K may be somewhat
lower. Alternatively, the efficiency $\eta$ of converting the
compressional energy release into the observed luminosity may be
lower. Within the uncertainties, however, it is also possible that the
present temperature is almost entirely due to compressional heating
and that the white dwarf had cooled to a temperature substantially
below 15\,000 K prior to the onset of mass transfer. In any case, the
cooling age of $1.3\times 10^8$\,yrs to reach 15\,000 K
\cite{wood95-1} is a lower limit to the actual pre-CV age of the white
dwarf. Since the Kelvin-Helmholtz time scale of the envelope is
roughly of the order of $10^4$\,yrs, the low temperature of the white
dwarf requires $\dot M$ to have been low for a comparable length of
time.

If RX\,J1313 is a CV in the process of turning on mass transfer
we would expect that the accretion rate in RX\,J1313 would ultimately
reach $\sim 10^{-9}$\,\msy\ at which time the white dwarf has been
compressionally heated to $\Twd\approx 30\,000$\,K, the temperature
typically observed in CVs with \Porb $\simeq 4$\,h \cite{sion99-1}.

\section{Conclusion}

We conclude that the low temperature of the white dwarf in RX\,J1313
is consistent with compressional heating by mass accretion at a rate
substantially lower than the \mbox{$\sim 10^{-9}$\,\msy} expected for
a long period CV. The system has not passed through a phase of high
accretion rate within at least the last $10^4$\,yrs which is the
approximate Kelvin-Helmholtz time scale for the envelope. There are
three possible previous histories of RX\,J1313: (a) it is a young CV
in the process of turning on the mass transfer; (b1) it is in a
long-lasting phase of low accretion within an irradiation-driven limit
cycle; (b2) it has passed through a nova outburst shutting off mass
transfer for a prolonged period.  We cannot presently distinguish
between cases (a) and (b1), while observational tests of (b2) have
been suggested above.

\begin{acknowledgements}
We thank Klaus Reinsch for the optical spectrum of RXJ\,1313 and for
useful comments on the manuscript, and Stefan Jordan for discussions
on magnetic model atmospheres. This research was supported by the DLR
under grant 50\,OR\,96\,09\,8 and 50\,OR\,99\,03\,6.
\end{acknowledgements}

\end{document}